\documentclass[12pt]{article}

\usepackage[dvips]{graphicx}

\def\dspace{\baselineskip = 0.30in}

\def\lapproxeq{\lower .7ex\hbox{$\;\stackrel{\textstyle
<}{\sim}\;$}}
\def\gapproxeq{\lower .7ex\hbox{$\;\stackrel{\textstyle
>}{\sim}\;$}}

\begin{document}

\dspace

\begin{titlepage}
\begin{flushright}
BA-03-03\\
\end{flushright}
\vskip 2cm
\begin{center}
{\Large\bf
Grand Unified Inflation Confronts WMAP 
}
\vskip 1cm
{\normalsize\bf
Bumseok Kyae\footnote{bkyae@bartol.udel.edu} and
Qaisar Shafi\footnote{shafi@bxclu.bartol.udel.edu}
}
\vskip 0.5cm
{\it Bartol Research Institute, University of Delaware, \\Newark,
DE~~19716,~~USA\\[0.1truecm]}

%

\end{center}
\vskip .5cm


\begin{abstract}
In a class of realistic four and five dimensional supersymmetric grand unified 
models, the scalar spectral index is found to be $n_s=0.98 (\pm 0.01)$, 
in excellent agreement with the values determined by several previous   
experiments and most recently by 
the Wilkinson Microwave Anisotropy Probe (WMAP).  The models predict 
$dn_s/d{\rm ln}k\sim 10^{-3}$ and a negligible tensor-to-scalar ratio 
$r\sim 10^{-8}$.  A new five dimensional supersymmetric $SO(10)$ model 
along these is presented in which inflation is associated with the breaking 
of $SO(10)$ to $SU(5)$ at scale $M$, 
with $\delta T/T\propto (M/M_{\rm Planck})^2$, 
so that $M\simeq 10^{16}$ GeV.   
The inflaton decay leads to the observed baryon asymmetry via leptogenesis.  
We also discuss how the monopole problem is solved without the use of 
non-renormalizable terms.  
  
\end{abstract}


\end{titlepage}

\newpage



Supersymmetric grand unified theories (GUTs) provide an especially attractive 
framework for physics beyond the standard model (and MSSM), and 
it is therefore natural to ask if there exists in this framework a compelling, 
perhaps even an intimate connection with inflation.  
In ref.~\cite{hybrid} one possible approach to this question was presented.  
In its simplest realization, inflation is associated with the breaking 
at scale $M$ of a grand unified gauge group $G$ to $H$.  Indeed, inflation is 
`driven' by quantum corrections which arise from the breaking of 
supersymmetry by the vacuum energy density in the early universe.  
The density fluctuations, it turns out, are proportional to 
$(M/M_{\rm Planck})^2$, where $M_{\rm Planck}\simeq 1.2\times 10^{19}$ GeV 
denotes the Planck mass.  From the variety of $\delta T/T$ measurements, 
especially by the Wilkinson Microwave Anisotropy Probe (WMAP)~\cite{wmap}, 
the symmetry breaking scale $M$ is of order $10^{16}$ GeV, 
essentially identical to the scale of supersymmetric grand unification.   

Because of the logarithmic radiative corrections that drive inflation, 
the spectrum of scalar density fluctuations turns out to be essentially flat.  
For the simplest models, the scalar spectral index was found to be 
$n_s=0.98(\pm 0.01)$~\cite{hybrid}, in excellent agreement
with a variety of observations~\cite{obs} including the recent WMAP data.  
The variation $dn_s/d{\rm ln}k$ of the spectral index is found to be small 
($\sim 10^{-3}$).  

In some recent papers~\cite{ks,ks2} it was shown 
how the above scheme can be extended to 
five dimensional supersymmetric models.  
There are good reasons for discussing such models.  Consider, for instance, 
the case of $G=SO(10)$ (or $SU(5)$) in four dimensions.  
The presence of dimension five baryon number violating operators 
mediated through Higgsino exchange implies 
in the `minimal' scheme a proton life time 
$\tau_{p\rightarrow K^+\bar{\nu}}\sim 10^{30\pm 2}$ yrs.  
This may be in conflict with the recent lower bounds 
($\tau_p> 1.9\times 10^{33}$ yrs) for $p\rightarrow K^+\bar{\nu}$ 
determined by the Superkamiokande experiment~\cite{kamio}.  
There are other serious issues such as the notorious doublet-triplet (DT) 
splitting problem, 
which have led people to investigate five (and higher) dimensional theories 
compactified on suitable orbifolds that provide a relatively painless way 
of implementing the DT splitting.  
Furthermore, dimension five proton decay can be easily eliminated which is     
an attractive feature of the five dimensional framework.  

In this paper we present a realistic model of inflation based on 
five dimensional supersymmetric $SO(10)$ compactified on an orbifold 
$S^1/(Z_2\times Z_2')$.  There are two fixed points (branes) where the gauge 
symmetries are $SO(10)$ and $SU(4)_c\times SU(2)_L\times SU(2)_R$,  
respectively~\cite{dermisek,ks}.  
Through the spontaneous breaking of $SO(10)$ to $SU(5)$, 
the effective low energy symmetry corresponds to the MSSM gauge group.  
The inflationary scenario will be associated with the symmetry breaking 
$SO(10)\rightarrow SU(5)$.    
In its simplest realization the scalar spectral index $n_s=0.98\pm 0.01$.  
The variation of $n_s$ with respect to the wave number $k$ is small 
but in principle measurable ($dn_s/d{\rm ln}k\sim 10^{-3}$), 
which will be tested by ongoing and future observations.  
The WMAP data combined with other observations appears to prefer, 
but does not require, a far more significant running, 
$dn_s/d{\rm ln}k=-0.031^{+0.016}_{-0.018}$.  
Our model predicts a negligible tensor-to-scalar ratio $r\sim 10^{-8}$.  
After inflation is over, the inflaton (which belongs to ${\bf 16}$, 
${\bf \overline{16}}$ of $SO(10)$) decays into right-handed neutrinos 
whose out of equilibrium leads to the observed baryon asymmetry 
via leptogenesis~\cite{lepto,lasymm,ks}.  
%
The symmetry breaking scale $M$ is estimated from inflation to 
be of order $10^{16}$ GeV, which leads to righ-handed neutrino masses 
of the correct magnitude, of order $10^{14}$ GeV or less, 
that can yield a mass spectrum 
for the light neutrinos suitable for neutrino oscillations~\cite{ks}.  

In this paper we also provide a new resolution of the well known monopole 
problem.  In the class of models discussed, unless care is exercised, 
superheavy monopoles can be produced at the end of inflation leading to 
cosmological disaster.  In ref.~\cite{khalil}, the problem was circumvented 
by including sizable higher order (non-renormalizable) terms 
in the superpotential, such that the GUT symmetry is broken 
along an inflationary trajectory.  
In our new scenario we show how the problem is solved 
in a five dimensional framework 
without invoking non-renormalizable terms.  

%
%

The four dimensional inflationary model is best illustrated by considering
the following superpotential which allows the breaking of a gauge
symmetry $G$ down to $H$,
keeping supersymmetry intact \cite{hybrid,lyth}:
\begin{eqnarray} \label{simplepot}
W_{\rm infl}=\kappa S(\phi\bar{\phi}-M^2) ~.
\end{eqnarray}
Here $\phi$ and $\bar{\phi}$ represent superfields
whose scalar components acquire non-zero vacuum expectation values (VEVs),    
which break $G$ to $H$.
The singlet superfield $S$ provides the scalar field that drives inflation.
Note that by invoking a suitable $R$ symmetry $U(1)_R$,
the form of $W$ is unique at the renormalizable level.   
For example, $W$ and $S$ can be assigned an $R$-charge of unity, 
while the $R$-charges of $\phi$, $\bar{\phi}$ are zero.   
It is gratifying to realize that $R$ symmetries naturally occur
in (higher dimensional) supersymmetric theories
and can be appropriately exploited.

From $W$, it is straightforward to show that the supersymmetric minimum
corresponds to non-zero (and equal in magnitude) VEVs
for $\phi$ and $\bar{\phi}$, while $\langle S\rangle =0$.  
(After supersymmetry breaking {\it $\grave{a}$ la} $N=1$ supergravity, 
$\langle S\rangle$ acquires a VEV
of order $m_{3/2}$ (gravitino mass)).

An inflationary scenario is realized in the early universe
with both $\phi$, $\bar{\phi}$ and $S$ displaced
from their present day minima.
Thus, for $S$ values in excess of the symmetry breaking scale $M$, 
$\phi$, $\bar{\phi}$ VEVs vanish,
the gauge symmetry is restored, and a potential energy density proportional
to $M^4$ dominates the universe. With supersymmetry thus broken, there are
radiative corrections from the $\phi$-$\bar{\phi}$ supermultiplets
that provide logarithmic corrections to the potential which drives inflation.
In one loop approximation \cite{hybrid,coleman},
\begin{eqnarray}\label{scalarpot}
V\simeq \kappa^2M^4\bigg[1+\frac{\kappa^2{\cal N}}{32\pi^2}\bigg(
2{\rm ln}\frac{\kappa^2|S|^2}{\Lambda^2}+(z+1)^2{\rm ln}(1+z^{-1})
+(z-1)^2{\rm ln}(1-z^{-1})\bigg)\bigg]~,
\end{eqnarray}
where $z=x^2=|S|^2/M^2$, ${\cal N}$ is the dimensionality of 
the representations to which $\phi$, $\bar{\phi}$ belong, and  
$\Lambda$ denotes a renormalization mass scale.
The logarithmic loop corrections in Eq.~(\ref{scalarpot}) enable  
the inflaton field to slowly roll down to the supersymmetric vacuum state.   
From Eq.~(\ref{scalarpot}) the microwave CMB anisotropy 
on the Hubble scale $l$ is found to be~\cite{hybrid}
\begin{eqnarray}\label{T}
\bigg(\frac{\delta T}{T}\bigg)_l\simeq \frac{8\pi}{\sqrt{\cal N}}
\bigg(\frac{N_l}{45}\bigg)^{1/2}\bigg(\frac{M}{M_{\rm Planck}}\bigg)^2
x_l^{-1}y_l^{-1}f(x_l^2)^{-1} ~.
\end{eqnarray}
Here, $y_l\simeq x_l(1-7/12x_l^2+\cdots)$, $f(x_l^2)^{-1}\simeq 1/x_l^2$,
for $S_l$ sufficiently larger than $M$,
and $N_l\simeq 50-60$ denotes the e-foldings needed to resolve the horizon
and flatness problems.

Comparison of the expression for $\delta T/T$ in Eq.~(\ref{T}) with the
WMAP result shows 
that the gauge symmetry breaking scale $M$ is    
around $10^{16}$ GeV~\cite{ns}, which is tantalizingly close to 
the GUT scale inferred from the evolution of the MSSM gauge couplings.   
Thus, it is natural to embed this kind of inflationary scenario 
within a GUT framework.   
However, in this case we must make sure that cosmological problems 
associated with topological defects such as monopoles do not arise. 
In addition, while constructing a realistic inflationary model based on 
supersymmetric GUT, we would also like to resolve the notorious DT splitting 
problem.  These, as we will see, are most easily carried out 
in a five dimensional framework.    

The scalar spectral index $n_s$ is approximately given by
\begin{eqnarray} \label{spectral}
n_s\simeq 1-\frac{1}{N}
\end{eqnarray}
where $N$ denotes the number of e-foldings experienced by the scale under
consideration.  For the horizon scale, in particular, $N_h\simeq 50-60$,
so that
\begin{eqnarray}
n_s^{(h)}= 0.98\pm 0.01 ~.
\end{eqnarray}
It should be noted that the inclusion of supergravity corrections
can, in some cases, lead to a spectral index larger than unity
[for a recent discusion and additional references, see ref.~\cite{ns}.].  
The galactic scale corresponds to $N_g\simeq 40-50$, 
and given Eq.~(\ref{spectral}), we conclude that the variation of
$n_s$ with $k$ is quite tiny, 
$dn_s/d{\rm ln}k\sim 10^{-3}-{\rm few}\times 10^{-4}$.
It would be interesting to test this prediction against the ongoing and future
observations.

As shown in refs.~\cite{khalil,hybrid2,ks}, a combination of the
the gravitino constraint on the reheat temperature
($T_R\leq 10^{10}$ GeV~\cite{gravitino})
%
%
as well as leptogenesis
requires that the dimensionless superpotential coupling $\kappa\sim 10^{-3}$.
Thus, the vacuum energy density during inflation is
of order $10^{-6}M_{\rm GUT}^4$, so that
the tensor-to-scalar ratio $r\sim 10^{-8}$,
which could be hard to detect in any forseeable experiment.

After inflation is over the universe converts to a radiation dominated epoch
through the superpotential couplings
$\gamma_{ij}\bar{\phi}\bar{\phi}{\bf 16}_i{\bf 16}_j/M_P$,
where ${\bf 16}_i$ ($i=1,2,3$) denote the three chiral families of $SO(10)$
(with $R$-charge $=1/2$), $\gamma_{ij}$ is a dimensionless coupling,
and $M_P$ ($\simeq 2.4\times 10^{18}$ GeV) is the reduced Planck mass.
That is, the inflaton decay produces right-handed neutrinos
whose out of equilibrium decay produces the observed baryon asymmetry via
leptogenesis along the lines previously discussed in~\cite{lepto,lasymm,ks}.

Next we present a realistic five dimensional $SO(10)$ model in which 
the inflationary scenario described by the superpotential $W$ 
in Eq.~(\ref{simplepot}) can be realized. 
We assume compactification on an orbifold $S^1/(Z_2\times Z_2')$,  
such that on the two fixed points (branes) we have the gauge symmetries
$SO(10)$ and $SU(4)_c\times SU(2)_L\times SU(2)_R$ respectively.  
To realize the MSSM gauge group at low energies, 
we introduce two pairs of the Higgs hypermultiplets ${\bf 16}_H$ and 
${\bf \overline{16}}_H$ in the bulk with $Z_2\times Z_2'$ parities, 
\begin{eqnarray}\label{1}
&&{\bf 16}_H~={\bf (\overline{4},1,2)}_H^{++}~+~{\bf (4,2,1)}_H^{+-} ~~,\\
&&{\bf 16^c}_H={\bf (\overline{4},1,2)}_H^{c--}+~{\bf (4,2,1)}_H^{c-+} ~,\\
&&{\bf \overline{16}}_H~   \label{3}
={\bf(4,1,2)}_H^{++}~+~{\bf (\overline{4},2,1)}_H^{+-}~~, \\
&&{\bf \overline{16}^c}_H
={\bf(4,1,2)}_H^{c--}+~{\bf (\overline{4},2,1)}_H^{c-+} ~.  
\end{eqnarray} 
The relevant superpotentials on the two branes, B1 ($SO(10)$ brane) and 
B2 ($SU(4)_c\times SU(2)_L\times SU(2)_R$ brane) are: 
\begin{eqnarray} 
&&W_{B1}=\kappa S\bigg({\bf 16}_H{\bf \overline{16}}_H-M_1^2\bigg) ~, \\
&&W_{B2}=\kappa S\bigg(c_1H^c\overline{H}^c
+c_2{\bf 1}{\bf 1'}
-M_2^2\bigg)+c_3\Sigma H^c\overline{H}^c ~,  \label{wb2}
\end{eqnarray}
where $H^c\equiv {\bf (\overline{4},1,2)}_H^{++}$, 
$\overline{H}^c\equiv {\bf(4,1,2)}_H^{++}$, and $c_1$, $c_2$, $c_3$ are 
dimensionless couplings.  
In $W_{B2}$, we exhibit only the chiral multiplets with $(++)$ parities  
of ${\bf 16}_H$, ${\bf \overline{16}}_H$ which contain massless modes, 
since the heavy KK modes would be decoupled.  
Since the inflaton $S$ is a bulk superfield,
it participates in both superpotentials.  
In Eq.~(\ref{wb2}), a pair of singlet superfields ${\bf 1}$, ${\bf 1'}$ and 
a superfield $\Sigma$ in the adjoint representation ${\bf (15,1,1)}$ 
with suitable $U(1)_R$ charges are introduced on B2.  

During inflation, $S$ and $\Sigma$ develop VEVs 
($\langle S\rangle>M_1,M_2$), 
while $\langle{\bf 16}_H\rangle=\langle{\bf \overline{16}}_H\rangle
=\langle H^c\rangle=\langle\overline{H}^c\rangle
=\langle{\bf 1}\rangle=\langle{\bf 1'}\rangle=0$.  
As shown in refs.~\cite{ks,ks2},
positive vacuum energies localized on the branes could trigger
exponential expansion of the three space, 
in the presence of a brane-localized Einstein-Hilbert term.  
Due to a non-zero VEV of $\Sigma$ during inflation, the $SU(4)_c$ factor 
in $SU(4)_c\times SU(2)_L\times SU(2)_R$ is spontaneously broken to 
$SU(3)_c\times U(1)_{B-L}$, and 
the accompanying monopoles are inflated away.     
%
%

%
%
In this brane model, ${\bf 16}_H$, ${\bf \overline{16}}_H$ on B1, and 
${\bf 1}$, ${\bf 1'}$ on B2 play the role of $\phi$, $\bar{\phi}$ 
in Eq.~(\ref{simplepot}).  
With the (localized) VEVs of the scalar components of ${\bf 16}_H$, 
${\bf \overline{16}}_H$ along the $SU(5)$ singlet direction 
(i.e. $\langle\nu^c_H\rangle$, $\langle\overline{\nu}^c_H\rangle$) at B1 
after inflation, the $SO(10)$ gauge symmetry breaks to $SU(5)$.   
On the other hand, at B2 only the singlets 
${\bf 1}$, ${\bf 1'}$ rather than 
$H^c$, $\overline{H}^c$ develop VEVs at the minimum of the potential.   
Since $\Sigma$ becomes heavy by VEVs of 
${\bf 16}_H$, ${\bf \overline{16}}_H$ on B1,      
the VEV $\langle\Sigma\rangle$ vanishes after inflation, and so  
the symmetry $SU(4)_c\times SU(2)_L\times SU(2)_R$ on B2 is restored.  
Consequently, the effective low energy theory after inflation  
is the desired MSSM  
($=\{SU(5)\}\cap \{SU(4)_c\times SU(2)_L\times SU(2)_R\}$).  
We note that the symmetry breaking process 
$SU(3)_c\times U(1)_{B-L}\times SU(2)_L\times SU(2)_R\rightarrow 
SU(3)_c\times SU(2)_L\times U(1)_{Y}$ does not create any unwanted 
topological defects such as monopoles, and so 
we have formulated a realistic 5D model in which the monopole problem 
is solved without introducing non-renormalizable terms.

%
%
%
%

While the Goldsotne fields ${\bf (\overline{3},1)}_{-2/3}^{++}$, 
${\bf (1,1)}_{-1}^{++}$ 
(also ${\bf (3,1)}_{2/3}^{++}$, ${\bf (1,1)}_{1}^{++}$) 
of $H^c$ ($\overline{H}^c$) are absorbed by the appropriate gauge bosons,      
the superhiggs mechanism leaves intact the massless supermultiplets 
${\bf (\overline{3},1)}_{1/3}$, ${\bf (3,1)}_{-1/3}$,       
which can acquire masses of order $m_{3/2}$ 
from their couplings to $\langle S\rangle$ after supersymmetry breaking.  
To eliminate this pair from the low energy theory,   
we can introduce on B1 a 10-plet with couplings
${\bf 16}_H{\bf 16}_H{\bf 10}$ and 
${\bf \overline{16}}_H{\bf \overline{16}}_H{\bf 10}$ (thus, 
${\bf 10}$ has an $R$-charge of unity), 
and/or a ${\bf (6,1,1)}$ field ($\equiv D$) on B2 with couplings  
$H^cH^cD$ and $\overline{H}^c\overline{H}^cD$.   
Then, the pair acquires superheavy masses proportional to 
$\langle\nu^c_H\rangle$ or $\langle\overline{\nu}^c_H\rangle$, and   
the low energy spectrum is precisely the MSSM one.  

Note that we introduced the Higgs 16-plets in the bulk rather than 
on the $SO(10)$ brane B1 in order to avoid unwanted states 
associated with the pseudo-Goldstone symmetry of the superpotential.   
Recall that the orbifold compactification breaks $SO(10)$ down 
to $SU(4)_c\times SU(2)_L\times SU(2)_R$.

%
%

To resolve the DT splitting problem, the Higgs 10-plet ( or ${\bf (1,2,2)}$) 
should be introduced 
in the bulk (on B2).  
By suitable $Z_2\times Z_2'$ parity assignments, the MSSM 
Higgs doublets are kept light, while the color triplets become superheavy.

In summary, we have taken the approach that a satisfactory inflationary 
scenario should:  

\noindent (i) resolve the flatness and horizon problem; 

\noindent (ii) resolve cosmological problems associated with topological 
defects;  

\noindent (iii) give rise to the observed $\delta T/T$ fluctuations; 

\noindent (iv) provide a satisfactory explanation of the origin 
of the observed baryon asymmetry; 

\noindent (v) be well grounded in particle physics.      

\noindent While four dimensional $SO(10)$ models of inflation are hard 
to construct, especially if a resolution of DT splitting problem is 
also desired, things are much easier if we consider five dimensional $SO(10)$.  
In the model we have discussed, the gauge symmetry during inflation is broken 
to $SU(3)_c\times U(1)_{B-L}\times SU(2)_L\times SU(2)_R$ so that monopoles 
are inflated away.  When inflation ends, the unbroken gauge symmetry turns 
out to be $SU(3)_c\times SU(2)_L\times U(1)_Y$.  
The scalar spectral index $n_s$ is very close (or even equal~\cite{ns}) 
to unity, $dn_s/d{\rm ln}k\sim 10^{-3}$, and the tensor-to-scalar ratio 
$r\sim 10^{-8}$.  The observed baryon asymmetry naturally follows 
from leptogenesis.  

%
%

\vskip 0.3cm
\noindent {\bf Acknowledgments}

\noindent
We thank Zurab Tavartkiladze and Yasunori Nomura for useful discussions.
%
The work is partially supported
by DOE under contract number DE-FG02-91ER40626.


\begin{thebibliography}{99}

\bibitem{hybrid} G. Dvali, Q. Shafi, and R. Schaefer,
Phys. Rev. Lett. {\bf 73}, 1886 (1994) [hep-ph/9406319].  
For a comprehensive review and additional references, see
G. Lazarides, hep-ph/0111328.  

\bibitem{cobe} 
G. F. Smoot {\it et. al.}, Astrophys. J. Lett. {\bf 396} L1 (1992); 
C. L. Bennett {\it et. al.}, Astrophys. J. Lett. {\bf 464}, 1 (1996).  

\bibitem{wmap} D. N. Spergel {\it et. al.}, astro-ph/0302209; 
C. L. Bennett {\it et. al.}, astro-ph/0302207; 
H. V. Peiris {\it et. al.}, astro-ph/0302225.  

\bibitem{obs} S. L. Bridle, A. M. Lewis, J. Weller, and G. Efstathiou, 
astro-ph/0302306;  
A. Lasenby, Talk presented at the 21st Texas Symposium 
On Relativistic Astrophysics,  Florence, Italy (Dec. 9--13, 2002).  

\bibitem{ks} B. Kyae, and Q. Shafi, Phys. Lett. {\bf B556}, 97 (2003) 
[hep-ph/0211059].  

\bibitem{ks2} B. Kyae, and Q. Shafi, to appear in  Phys. Rev. {\bf D} 
[hep-ph/0212331].  

\bibitem{kamio} For example, Y. Totsuka, Talk at SUSY 2K, CERN, June (2000).  

\bibitem{dermisek} R. Dermisek, and A. Mafi, Phys. Rev. {\bf D65},
055002 (2002) [hep-ph/0108139]; 
C. H. Albright, and S. M. Barr, Phys. Rev. {\bf D67}, 013002 (2003) 
[hep-ph/0209173]; 
H. D. Kim, and S. Raby; JHEP {\bf 0301}, 056 (2003) [hep-ph/0212348].  

\bibitem{lepto} M. Fukugita, and T. Yanagida,
Phys. Lett. {\bf B174}, 45 (1986); 
G. Lazarides, and Q. Shafi, Phys. Lett. {\bf B258}, 305 (1991); 
L. Covi, E. Roulet, and F. Vissani, Phys. Lett. {\bf B384}, 169 (1996) 
[hep-ph/9605319]. 

\bibitem{lasymm} G. Lazarides, Q. Shafi, and N. D. Vlachos, Phys. Lett.
{\bf B427}, 53 (1998), and references therein.

\bibitem{khalil} R. Jeannerot, S. Khalil, G. Lazarides, and Q. Shafi,
JHEP {\bf 0010}, 012 (2000) [hep-ph/0002151].


\bibitem{lyth} E. J. Copeland, A. R. Liddle, D. H. Lyth, E. D. Stewart, and
D. Wands, Phys. Rev. {\bf D49}, 6410 (1994).

\bibitem{coleman} S. Coleman and E. Weinberg,
Phys. Rev. {\bf D7}, 1888 (1973). 

%
%

\bibitem{ns} V. N. Senoguz, and Q. Shafi, hep-ph/0305089. 

%
%

\bibitem{hybrid2} G. Lazarides, R. K. Schaefer, and Q. Shafi,
Phys. Rev. {\bf D56}, 1324 (1997) [hep-ph/9608256].

\bibitem{gravitino} J. Ellis, J. E. Kim, and D. Nanopoulos,
Phys. Lett. {\bf B145}, 181 (1984);
M. Yu. Khlopov, and A. D. Linde, Phys. Lett. {\bf B138},
265 (1984). For a review and additional references,
see W. Buchm${\rm \ddot{u}}$ller,
Nato Science Series II, Vol. 34, 2001, eds. G. C. Branco, Q. Shafi, and
J. I. Silva-Marcos.

%
%

\end{thebibliography}
\end{document}